\documentclass{article}
\usepackage[T1]{fontenc}
\usepackage{wrapfig}
\usepackage[portuges,english]{babel}
\usepackage{amssymb,latexsym,amsmath,color,mathrsfs,ifsym,graphics,stmaryrd} 
\usepackage[colorlinks,linkcolor=blue,urlcolor=blue,citecolor=black,
plainpages=false,pdfpagelabels,breaklinks]{hyperref}
\usepackage[all]{xy}
\usepackage{graphicx}

\title{{\sc The Dilemma of Quantum Individuality\\Beyond Particle Metaphysics}}

\author{{\sc Christian de Ronde}$^{1,2, 3, 4, 5}$  and {\sc Raimundo Fern\'andez-Mouj\'an}$^{1, 5}$}
\date{}

\usepackage[margin=2.2cm]{geometry}

\begin{document}

\bibliographystyle{plain}
\maketitle

\begin{center}
\begin{small}
1. University of Buenos Aires\\
2. Philosophy Institute Dr. A. Korn, UBA-CONICET\\
3. National University Arturo Jauretche, Argentina.\\
4. Federal University of Santa Catarina, Brasil.\\
5. Center Leo Apostel for Interdisciplinary Studies, Vrije Universiteit Brussel, Belgium.\\
\end{small}
\end{center}

\begin{abstract}
\noindent It is commonly claimed that quantum mechanics makes reference to a microscopic realm constituted by elementary particles. However, as first famously noticed by Erwin Schr\"odinger, it is not at all clear what these quantum particles really are. According to the specialized literature, it is not even clear if each of these microscopic entities possess their own {\it identity}. Recently, Jonas Arenhart proposed a distinction of quantum objects in terms of a dilemma which forces a choice between their characterization either as {\it individuals} or as {\it non-individuals}. In this work we attempt to address the (metaphysical) presuppositions involved within Arenhart's dilemma which ground the question of individuality in QM on a strong presupposition regarding the existence of quantum objects. After providing a reconsideration of the role played by metaphysics within physics we attempt to propose, not only a complete redefinition of the dilemma beyond particle metaphysics, but also a possible realist solution grounded on the provision of a new (non-classical) conceptual framework which seeks to develop an invariant-objective representation of the theory of quanta. 
\end{abstract}
\begin{small}

\end{small}

\newtheorem{theo}{Theorem}[section]
\newtheorem{definition}[theo]{Definition}
\newtheorem{lem}[theo]{Lemma}
\newtheorem{met}[theo]{Method}
\newtheorem{prop}[theo]{Proposition}
\newtheorem{coro}[theo]{Corollary}
\newtheorem{exam}[theo]{Example}
\newtheorem{rema}[theo]{Remark}{\hspace*{4mm}}
\newtheorem{example}[theo]{Example}
\newcommand{\proof}{\noindent {\em Proof:\/}{\hspace*{4mm}}}
\newcommand{\qed}{\hfill$\Box$}
\newcommand{\ninv}{\mathord{\sim}} 
\newtheorem{postulate}[theo]{Postulate}

\bigskip

\bigskip

\bigskip

\bigskip

\bigskip

\begin{flushright}
{\it We are, to be sure, all of us aware of the situation\\ 
regarding what will turn out to be the basic\\ 
foundational concepts in physics: the point-mass\\ 
or the particle is surely not among them.}\\
\smallskip
\smallskip

Albert Einstein. 
\end{flushright}

\section{What is an Elementary Particle?}

It is difficult to overstate the influence of atomist philosophy in modern and contemporary physical science. Since Newton's development of classical mechanics in the 17th Century, atomism became the ``common sense'' perspective from which physicists regraded the world. And even though this understanding was shaken in the 19th Century, due to the appearance of the notion of `field' in Maxwell's theory of electromagnetism, during the 20th Century atomism returned with even more strength than before providing the missing picture of a strange theory that would soon be known under the name of Quantum Mechanics (QM). In this context of analysis, it is maybe useful to recall that atomism started already two and a half millennia ago, when Leucippus and Democritus developed a philosophy that postulated the world as originated in two main elements: being and not-being, or the full and the void. Being was to be considered, by atomists, as tiny indivisible bodies, homogeneous in nature, but different in shape and size. These indivisible bodies moved around (continuously) in void, and according to their different combinations, they produced the totality of our very strange and rich reality. It is important to notice that this philosophy represents a substantivalist worldview, that is, an understanding of the world as made of multiple separated individual substances, a worldview that starts by independent substantial individuals and constructs the totality of reality by adding up these individuals. In modern science it was specially the development of classical mechanics which took this atomistic worldview and most definitively imprinted it in the mind of physicists. Newton was able to mathematize atoms in space as points in a {\it continuous} mathematical representation provided via the newly developed infinitesimal calculus. Through an equation of motion it was then possible to determine the evolution of physical systems as constituted by atoms. The picture of the world described by Newtonian mechanics was that of small perfectly determined particles bouncing against each other in space and in time. As remarked by Laplace, for anyone capable of knowing the exact position of all particles in the Universe, ``nothing would be uncertain and the future just like the past would be present before its eyes.'' 

But atomism is not the only philosophy that established a fundamental relation with the basic concepts of physical thought. We can also cite Aristotle's logic and ontology, whose principles allowed for a glimpse of permanence and determination in the thought about the physical world, that most Greek thinkers saw as always changing, undetermined and in constant becoming. Aristotle's development in regards to the formal (and actual) aspect of physical entities made it possible to develop a scientific discourse about sensible objects. The Principle of Non-Contradiction, or the Principle of Identity, for instance, played a key role, creating a representation of entities (as possessing identity, as being non-contradictory), and a discourse about them, that seemed in agreement with the possibility of true knowledge about the physical world. It is well known that Newton's physics followed the path paved by Aristotle's logical and ontological principles, but, choosing to erase the potential way of being that was also present within Aristotle's physics and metaphysics ---which accounted for the element of indetermination of entities--- Newton transformed particles in purely actual existents. Classical physics, in the 17th Century, mainly due to its enormous representational and predictive power, became the frame for all thought and discourse about physical reality ---now understood in terms of Aristotle's actual mode of existence. After the creation of classical mechanics, another philosophical work that had a decisive role in the development of modern physical thought is that of Kant. The Kantian {\it transcendental subject} allowed to save science from the skeptical wave that menaced at the time to discard the possibility of physical knowledge itself. He developed, in the {\it Critic of Pure Reason}, a new type of objective knowledge grounded on transcendental conditions. This new objectivity not only saved natural science from the attacks of skeptics, it also allowed to legitimize and incorporate as a kernel basis of the new architectonic both the Aristotelian {\it categories} as well as the continuous {\it topos} introduced by Newtonian physics (i.e. absolute space and time) now understood as {\it forms of intuition} of the transcendental subject. These philosophical developments allowed for the articulation of an objective representation of the physical world as constructed of non-contradictory, separated objects in space and time (or, more contemporary, in space-time), that are ultimately build of really tiny (but still non-contradictory, separated, etc.) objects, which are the limit to the division of matter. 
But the influence of atomism did not end there. It continued to play an essential role in the development of physics in particular, and science in general. As pointed out by Werner Heisenberg: 
\begin{quotation}
\noindent {\small ``The strongest influence on the physics and chemistry of the past [19th] century undoubtedly came from the atomism of Democritos. This view allows an intuitive description of chemical processes on a small scale. Atoms can be compared with the mass points of Newtonian mechanics, and from this a satisfactory statistical theory of heat was developed. [...] the electron, the proton, and possibly the neutron could, it seemed, be considered as the genuine atoms, the indivisible building blocks, of matter. In this way the atomism of Democritos became an essential part of the materialistic interpretation of the world during the past century: easily understood and intuitively plausible, it determined the way of thinking of even those physicists who insisted on not dealing with philosophy.'' \cite[p. 218]{Castellani98}}
\end{quotation}

\noindent Ernst Mach, one of the main figures of 19th Century positivism, had strongly criticized the (metaphysical) postulation of atoms. But, quite paradoxically, his positivist philosophy ---specially his critical analysis of classical mechanics--- allowed the young generation of physicists to develop a new \textit{atomic} theory grounded on {\it quanta}. During the 20th Century, QM developed from the interrelation of two contradictory standpoints. The first was atomist metaphysics, which played an essential role in the reference of the theory to elementary quantum particles. The second ---maybe even more important--- was the empirical-positivist (anti-metaphysical) understanding of science as making reference to observations. As it is evident, these two stances are incompatible: while atomist philosophy postulates strong metaphysical thesis, positivism was quite explicitly developed from a mistrust in metaphysical thought and represented an attempt to clear metaphysics out of its fundamental role in the development of scientific theories. But even though QM was born from the explicit application of positivist ideas and a radical departure from classical notions, this did not stop physicists from claiming anyhow that the mathematical formalism of the theory referred to ``elementary particles'' (i.e., atoms). Very soon, however, it became evident that these new quantum particles had very strange behaviors, not at all compatible with what one would expect from any classical object. Since the development of the mathematical formalism in the mid 1920s, first by Heisenberg and then by Erwin Schr\"odinger, it became very clear that the reference of the new theory departed drastically from the classical atomist representation of reality. Regardless of the first efforts to interpret the quantum wave function, $\Psi$, as a kind of ``material wave'' ---following the ideas of Louis de Broglie--- it soon became evident, to Schr\"odinger himself, that this was a dead end. $\Psi$ was mathematically represented in configuration space, which changed its dimension when considering a different number of quantum wave functions. In 1926, Max Born interpreted $\Psi$ as a strange kind of ``probability wave'' which could even interact with other wave functions of its class. But regardless of the fact that $\Psi$ was clearly not a wave nor a particle, the reference to elementary particles remained behind the scenes. In Born's probabilistic interpretation, the quantum wave function referred to the probability of finding a particle in a specific region of space ---this was done regardless of the fact the theory did not provide any representation of the mentioned particle. In tune with the positivist {\it Zeitgeist}, QM had abandoned its interest in the metaphysical representation of unobserved states of affairs and focused on the pragmatic predictive account ---if only probabilistically--- of observations and measurement outcomes. The reference beyond measurements seemed to be just an uninmportant ``way of talking'' without any consistency constraint. QM talked about the `clicks' in detectors produced by supposed \emph{elementary particles} and allowed to compute probabilistically the possibility of finding a particle (i.e., a `click') in a definite position. But what about the particle? Was the particle in that position before the measurement outcome had been observed? Bohr would explicitly discard this question, as he argued that it made no sense in QM, for we had reached a limit in our representation of nature. Werner Heisenberg  \cite[p. 42]{Heis58} would address this problem and, going back to the hylemorphic Aristotelian scheme, would claim that a quantum particle was ``something standing in the middle between the idea of an event and the actual event, a strange kind of physical reality just in the middle between possibility and reality.'' 

And as if that wasn't enough, the problems with the notion of particle and, in general, with the substantivalist representation that gave sense to the classical notion of object, arose even more dramatically with the new quantum statistics derived from the theory. Indeed, QM prescribed a new way of counting electrons, photons and the like. In opposition to classical systems, which obey Maxwell-Boltzmann statistics, quantum particles obey Bose-Einstein and Fermi-Dirac statistics. Consider two elementary particles such as, for example, electrons. If these particles were assumed to exist as distinct individuals, they could be obviously labeled as `electron 1' and `electron 2'. Suppose now we would like to placed them in two boxes. If they were objects, due to their individual existence in space-time, each electron could be placed either in the left box or in the right box. Thus, there would exist four different possibilities: both electrons in the left box, both in the right box, electron 1 in the left box and electron 2 in the right box, and viceversa, electron 2 in the left box and electron 1 in the right box. These `labeled' electrons would obey Maxwell-Boltzmann statistics and all four possibilities should be considered on equal footing. Each possibility would be then quantified with a probability of 0.25. So even if we could not epistemically distinguish the two electrons, the last two possibilities would sum up to a 0.5 probability. On the contrary, quantum statistics impose that the two last options should be quantified not only as a single possibility but also on equal footing to the first two possibilities. If the particles were bosons, there would only be three possibilities left ---each quantified with a probability of 0.33--- while if they were fermions, there would be only one possibility ---because, in addition, the two first arrangements are prohibited for fermions by the Pauli exclusion principle. It is clearly difficult to justify this strange way of counting the possible arrangements if particles are assumed to be individual objects existing in space-time. But instead of drawing the obvious conclusion, namely, that QM was not talking about particles, physicists followed a much more difficult path to justify. Physicists begun to argue that quantum particles had to be considered as ontologically indistinguishable, and consequently, could not be regarded as individuals. Thus, there were still quantum \emph{particles}, we could still talk about particles, but they simply lacked identity. 

In his famous paper: {\it What is an elementary particle?} Erwin Schr\"odinger wrote: ``Atomism in its latest form is called quantum mechanics. [...] In the present form of the theory the `atoms' are electrons, protons, photons, mesons, etc. The generic name is elementary particle, or merely particle. The term `atom' has (...) been retained for chemical atoms, though \textit{it has become a misnomer}.'' In fact, the difficulties in applying the notion of `particle' in QM had become already evident during the mid 1930s due to Einstein's and Schr\"odinger's critical analysis of the consequences implied by the mathematical formalism (see for a detailed discussion \cite{deRondeMassri19a}). In general, this critical questioning regarding the meaning of the theory was silenced by Bohr's triumph in the famous EPR battle. Bohr argued not only that QM had found a limit to our representation of the physical world, but also that, the theory was referring to a microscopic realm which could be understood in some kind of limit as paving the way to the classical one ---something that was stated in his typical ambiguous way in terms of a {\it correspondence principle} (see \cite{Boku14}). So Bohr played a major role not only in evacuating the need for a critical analysis of the atomistic metaphysics and language being assumed (this task had no sense since there was a limit to our knowledge), but also in creating the basic assumption that we are talking ---even if we can't say much about it--- about a microscopic, atomic realm that must be there since it is the base from where our classical, intuitively accepted reality comes from. Bohr's inconsistent reading was sedimented in Dirac's positivist formulation of QM in 1930. Very soon, with the approval of Bohr, his book {\it The Principles of Quantum Mechanics} \cite{Dirac74} together with the work of von Neumann \cite{VN} became known as the orthodox textbook formulation of the theory of quanta. Since then, every student has been taught that QM provides the correct (probabilistic) predictions to the occurrence of measurement outcomes but no representation of the entities that are supposedly responsible for producing those events. In his book, assuming a positivist standpoint, Dirac had stressed explicitly a warning to the attentive reader that justified this representational lacuna:  ``[it is]  important to remember that science is concerned only with observable things'', and also, that ``the main object of physical science is not the provision of pictures, but the formulation of laws governing phenomena and the application of these laws to the discovery of phenomena. If a picture exists, so much the better; but whether a picture exists of not is a matter of only secondary importance.'' Thus, regardless of the fact that there was no clear reference of the mathematical formalism of the theory, the reference to quantum particles remained as a necessary constituent of the quantum discourse. It is in this strange manner, and quite paradoxically, that atomism was saved in QM by its worst enemy. The positivist {\it Zeitgeist} of the 20th Century, which, as we said, was in fact based on the very denial of metaphysical representations kept using atomist philosophy as an unavoidable ``common sense'' discourse. Positivism didn't actually evacuate metaphysics from it's fundamental role in the constitution of physical theories, but instead crystalized one metaphysics in particular as an unavoidable given. Thus, while atomism was accepted without critical analysis, as a mandatory starting point ---with presuppositions that of course have consequences on what one is able to conceive---, positivism then proceeded to leave behind all other metaphysical endeavors considered now as storytellings about that which could not be observed.

\section{The Dilemma of Quantum Objects: Individuals or Non-Individuals?}

Very recently, Jonas Arenhart has proposed a characterization of the problem surrounding quantum individuality in terms of a dilemma which centers its attention in the notions of identity and indistinguishability. 
\begin{quotation}
\noindent {\small ``The Received View on quantum non-individuality (RV) is, roughly speaking, the view according to which quantum objects are not individuals. It seems clear that the RV finds its standard expression nowadays through the use of the formal apparatuses of non-reflexive logics, mainly quasi-set theory. In such logics, the relation of identity is restricted, so that it does not apply for terms denoting quantum particles; this `lack of identity' formally characterizes their non-individuality. We face then a dilemma: on the one hand, identity seems too important to be given up, on the other hand the RV seems to require that identity be given up.'' \cite[p. 1323]{Arenhart17}}
\end{quotation}

The idea that identity is fundamental for physical theories has been supported by several authors. Most recently, Ot\'avio Bueno has provided a set of arguments which attempt to secure the reference of theories to such notion \cite{Bueno14, Bueno17}. Against Bueno, the fundamentality of identity has been attacked by D\'ecio Krause and Arenhart \cite{KrauseArenhart19} who, on the contrary, have argued that the notion of ``identity is, for all practical purposes, unnecessary.'' Arenhart and Krause conclude that ``the arguments seeking to establish that identity is fundamental, according to Bueno, are unsuccessful. Almost every claim made to establish this thesis can be either shown not to achieve its goal or else to be amenable to be paraphrased in terms of discernibility or an alternative notion that does not involve identity. So, in the end, it seems that what we really need is at most a discernibility relation, which is in fact closer to our everyday necessities.'' The name given to the quantum objects which are indistinguishable but have no identity is that of `non-individual'. The idea that non-individuals are a meaningful notion capable of addressing quantum particles has been defended by Krause and Arenhart in several works \cite{ArenhartKrause14a, ArenhartKrause14b, KrauseArenhart19}. The main argument supporting the sense of non-individuals is provided through the reference to logical systems which show a limit to identity, namely, different type of non-reflexive logics \cite{daCostadeRonde14} and the quasi-set theory pioneered by Krause \cite{Krause92}. Following this path, one can think of indiscernibility as replacing identity. 
\begin{quotation}
\noindent {\small ``assume indiscernibility as a primitive term and recognize that it does not collapse with identity. We believe that the fact that indiscernibility can be analyzed without necessarily implying identity in some systems of logic shows that there is not a necessary equivalence between these notions: the fact that two items are indiscernible does not imply that they are identical. At the very least, it is logically possible that the relations of discernibility and difference are not the same, with discernibility being a weaker notion. In this case, there is an alternative way to understand the situation envisaged by Bueno without necessarily using identity. If this is correct, then identity is not really fundamental in the sense of being required for the meaningful application of concepts. The transcendental appeal of the argument goes by the board.''  \cite{KrauseArenhart19}}
\end{quotation}

Arenhardt and Krause aim at describing the curious nature of quantum objects by claiming that these are a particular type of objects to which identity simply does not apply. But what is the notion of `object' they suppose as primitive, and from which they depart to define this different kind of objects, the quantum ones? We can deduce, for instance, that identity would be one of its features, since it was part of it before it was subtracted to create the `quantum object'. This leads us to believe that the notion of object that is being used is the classical one we described at the beginning of our work, and that is in fact commonly presupposed. The notion of a `quantum object' seems to be defined then in negative terms, by subtracting aspects to the notion of object. And thus the term for it is a negative one: non-individual. But what is an object, understood classically, that is not an individual, that lacks one of its essential features (as postulated by Aristotle), namely identity? It seems it would be better to abandon the notion of object altogether. Arenhardt, together with Krause, claim that this should not be done, for the loss of this notion might imply the loss of rational discourse itself.
\begin{quotation}
\noindent {\small ``Quantum mechanics, on its fundamental level, is seen as objectless. But are we really allowed to draw such conclusions? What are the consequences of eliminating objects? Don't we speak about objects (mainly `particles''), describe objects, experiment on objects, and so on? Furthermore, doesn't the elimination of objects entail the end of rational discourse? As Muller [28] raises this issue: by abolishing objects aren't our traditional canons of inference also gone away?'' \cite{KrauseArenhart14}}
\end{quotation}
Instead, they propose a notion of `\emph{quantum} object' defined by subtracting to the object one of its main characteristics. But of course, the problem of intelligibility arises: how to represent an object that lacks identity? How to think of a non-individual? By subtracting characteristics to the notion of object, aren't we making it unintelligible? Arenhardt and Krause claim that it is possible to provide intelligibility through the provision of a logical formalism that would be capable to escape the reference to identity. The analysis then rapidly shifts to that of non-reflexive logics and quasi-set theory. But is it a logical formalism that which is able to provide intelligibility or rather a conceptual framework? Let's leave the answer to this question for a bit later, because it seems that the notion of object finds some other obstacles besides the `lack of identity' to be applied to QM. The theory poses a problem not only with identity, but actually with almost every principle that, as we explained earlier, determines the notion of object. While quantum superpositions, as shown clearly by Schr\"oinger with his zombie cat, shows a conflict with the principle of non-contradiction \cite{daCostadeRonde13}; entanglement, as shown once and again by Einstein, entails a problem with separability \cite{Howard10}. Even more problematic is the fact that QM faces serious difficulties in order to make sense of space: Planck's quantum postulate imposes a discrete type representation and $\Psi$ is described in a multi-dimensional {\it configuration space}. It is far more than identity what must be subtracted if we insist on taking as a starting point a presupposed notion of object. It seems we are left with very little, almost nothing, and no new logical formalism can make this mostly negative `quantum object' intelligible. 

But our conclusions should not be skeptical, since it is not true that we lose rational discourse when we reject the classical notion of object. Since the metaphysics that gives the notion of classical object its sense is directly presupposed, directly taken as ``common sense'', as reality itself, it should not surprise us that when this representation faces definitive obstacles, as it does with QM, the reaction is to fear the abyss of irrationality, the loss of scientific discourse. As a matter of fact, things are actually not so terrible nor dangerous once we are able to critically analyze the metaphysics behind classical objectivity, instead of simply presupposing it. The dilemma is not between a classical substantivalist worldview, presupposed as evident, and irrationality. The great problem with the enormous influence positivism had on physics and philosophy of science is that, by assuming that we could do without metaphysics, that all we needed was experimental data and its translation into a coherent mathematical formalism, metaphysics ---which in fact, no matter the intention, can't be evacuated from the fundaments of any scientific theory---, or rather a particular metaphysics, was directly and unconsciously presupposed in the sense given to what is observed: we see a spot in a photographic plate and we say a `particle' has been observed. It is not so difficult to see the unjustified deduction here. But, if we follow a different path, together with a critical analysis, we believe it is possible to advance in the development of a conceptual scheme in accordance to the quantum formalism. Such a conceptual development should take the different aspects of QM not as negative features, not as something that is lacking, not from the point of view of the metaphysics of classical physics, but positively, as the representation that should be captured by the theory itself.

In any case, Arenhardt and Krause, as well as Bueno, presuppose the same ground for understanding QM. Regardless of their differences they all seem to agree that QM talks about objects. Why? Not because the notion is particularly well suited for the quantum formalism, but, in the end, mainly because physicists talk in this way about the theory. This presupposes a specific understanding of physical theories as disconnected from their ontological reference. As in fact remarked in a recent paper co-authored by all three of them: ``Ontology is concerned with what exists and with what kinds of things exist. Although this description may sound abstract and far from the concerns of physics, the relation between ontology and physics is a close one. \emph{Of course, we are not claiming that physics cannot be successful without ontology. If that were the case, ontology would be required for physics, and it is not}.'' Physicists talk about electrons but they do not need to explain what they really are. That is a metaphysical or ontological task ---which does not affect `successful' physical theories---, and consequently, one that must be undertaken by the philosopher. As explained by Charles Sebens: 
\begin{quotation}
\noindent {\small ``Today, philosophers who are interested in figuring out what everything is made of look to contemporary physics for answers. But, finding answers in physics is not simply a matter of reading textbooks. Physicists deftly shift between different pictures of reality as it suits the task at hand. The textbooks are written to teach you how to use the mathematical tools of physics most effectively, not to tell you what things the equations are describing. It takes hard work to distill a story about what's really happening in nature from the mathematics. This kind of research is considered `philosophy of physics'.'' \cite{Chip19}}
\end{quotation}
This widespread understanding of physics ---which goes back to empirical positivism--- conceives theories as empirically adequate formal schemes capable of providing ---through a specific set of rules--- the prediction of observable measurement outcomes. What is of outmost important to stress is that, according to this scheme, the reference to an interpretation is something added {\it a posteriori} to an already empirically adequate ``successful'' theory, a supplementary discourse that does not have an influence on what is essential to the theory. So in the end it is not so terrible if we just make use of a common ``way of talking''. As remarked by Arenhart, Bueno and Krause: ``Physicists need not be concerned with ontological problems raised by physics, just as one need not be familiar with the Peano axioms in order to be able to use arithmetical operations.'' From this viewpoint, ``ontology is part of the enterprise, shared by most physicists [but not all], of obtaining information about how the world works and what it is made of.'' Such questions are not part of physics itself. Since metaphysics is just storytelling, and it entails no consequences regarding the ``success'' of theories, there are no fundamental risks involved in our choosing between these auxiliary discourses. But, as we explained, things are not so simple. To dismiss metaphysics as unimportant does not imply that the worldview that informs physical thought ---from way before QM appeared--- has, magically, no longer effect on us. It is fundamental, whether we want it or not. And the proof that it functions still, is our continuous inability to provide a representation of the quantum theory. Our substantivalist, atomist view does not represent just ``a way of talking'' with no influence over our thought, it is part of a worldview that informs \emph{a priori} our understanding of the physical world in general. So, not only metaphysics is important for science, but it is difficult to conceive a way forwards in our understanding of the reality expressed in our best physical theory without a critical analysis of our latent metaphysical views. These views are in conflict with QM, and this fact needs to be fully assumed. 

\smallskip 

Contemporary quantum physicists and philosophers of physics keep making reference ---either explicitly or implicitly--- to elementary quantum particles which are not described by QM. We fear that this way of talking is related more to an epistemological obstruction \cite{deRondeBontems11} ---physicists and philosophers have been unable to overcome--- than to a meaningful presentation of the theory. In the following section we attempt to discuss a reformulation of Arenhart's dilemma from a completely different (non-empiricist) standpoint which understands the role of metaphysics as truly essential within the development of physical theories.

\section{The Role of Metaphysics in Quantum Mechanics}

The reference to the ``way of talking'' by a scientific community has become one of today's main concerns for philosophy of science in order to address the foundation of theories. The claim is that theories talk about what scientist say they do. This justification of theories related to scientific practice and inter-subjective rationality became popularized by Thomas Kuhn, Paul Feyerabend and many others, during the 1960s and 1970s. As remarked by Helen Longino: ``Scientific rationality was to be understood by studying actual episodes in the history of science, not by formal analyses developed from {\it a priori} concepts of knowledge and reason (Kuhn 1962, 1977).'' Since the creation of the ``Big science'' the importance of scientific communities has become increasingly dominant. An explicit account of the centrality of what scientists actually say is reflected in Quine's naturalistic realism which, resumed by Arenhart  \cite[p. 13]{Arenhart19}, claims the following: ``There exists precisely what our best scientific theories say there is (see Quine 1981, p.21). There may be some room for disagreement here and there, but the world consists, basically, of the posits of contemporary science.'' Such an understanding of physical theories, as justified through the intersubjective communication between the members of a scientific community, implies in fact an explicit relativist standpoint. One that shifts from the subjective relativism of individuals to an inter-subjective relativism of communities. In this respect, `naturalistic realism' might seem as  a way of legitimizing relativism, a way of disguising it as scientific by extending the fallacy that what is said can be judged right or wrong on the basis of who is saying it. Science becomes a question of which is the best story, this is, which is the one most popular among scientists at the moment. But how can I know that `what is said' is a good criterion? How do I know that it results from aiming at true knowledge and not at, for instance, mere professional adaptation? How do I know if I’m not simply legitimizing prejudgments, assuming the truth of what are in fact preconceptions common to a group of professionals? How do I know if what I have in my hands is not the mere expression of a specific group of people or simply the mentality of a time? And these habits of thought, aren't they many times the main obstacles to our understanding of the world? Don't we have enough examples of that throughout history? This relativism is surely common in many of the trends of thought that became dominant throughout the 20th Century, not only in physics but also in philosophy. As explained by David Deutsch, one of the most critical voices against the contemporary empirical-positivist understanding of physical theories: 
\begin{quotation}
\noindent {\small ``[T]wentieth century, anti-realism became almost universal among philosophers, and common among scientists. Some denied that the physical world exists at all, and most felt obliged to admit that, even if it does, science has no access to it. [...] During the second half of the twentieth century, mainstream philosophy lost contact with, and interest in, trying to understand science as it was actually being done, or how it should be done. Following Wittgenstein, the predominant school of philosophy for a while was `linguistic philosophy', whose defining tenet was that what seem to be philosophical problems are actually just puzzles about how words are used in everyday life, and that philosophers can meaningfully study only that. [...] One currently influential philosophical movement goes under various names such as postmodernism, deconstructionism and structuralism, depending on historical details that are unimportant here. It claims that because all ideas, including scientific theories, are conjectural and impossible to justify, they are essentially arbitrary: they are no more than stories, known in this context as `narratives'. Mixing extreme cultural relativism with other forms of anti-realism, it regards objective truth and falsity, as well as reality and knowledge of reality, as mere conventional forms of words that stand for an idea's being endorsed by a designated group of people such as an elite or consensus, or by a fashion or other arbitrary authority. And it regards science and the Enlightenment as no more than one such fashion, and the objective knowledge claimed by science as an arrogant cultural conceit. Perhaps inevitably, these charges are true of postmodernism itself: it is a narrative that resists rational criticism or improvement, precisely because it rejects all criticism as mere narrative. Creating a successful postmodernist theory is indeed purely a matter of meeting the criteria of the postmodernist community ---which have evolved to be complex, exclusive and authority-based.'' \cite[pp. 313-314]{Deutsch04}}
\end{quotation}

Of course, since there is no systematic connection between physical theories and conceptual arrangements, there are many different narratives that can be created for exactly the same empirically adequate theory. This is known in the literature as metaphysical underdetermination.\footnote{Which adds to the well known underdetermination of the theory by experience, namely, the claim there are many possible mathematical formalisms capable to account for the same set of data.} As explained by Lawrence Sklar: ``Our foundational theories usually exist in a scientific framework in which they are subject to multiple, apparently incompatible, interpretations. And given the interpretation you pick, your view of what the theory is telling us about the basic structure of the world can be radically unlike that of someone who opts for a different interpretation of the theory.'' Obviously, for a (naive) scientific realist who wants to believe that her interpretation is true, this becomes an important obstacle. She needs to choose only one story between the many different ones without possessing any objective rule in order to make this difficult selection. Metaphysical underdetermination finds its most ridiculous expression in QM, where the number of narratives continues to grow with every passing year. Many worlds, many minds, decoherence, complementarity, flashes, information, histories, potentialities, propensities, latencies and even perspectivalism, they are all regarded by philosophers as reasonable candidate narratives for the theory of quanta. Today, each community fights for its own aspirant but consensus still looks like a distant chimera. This has led Adan Cabello to characterize the many interpretations of QM as a ``map of madness'' \cite{Cabello17}. However, regardless of their differences, and going back to our original problem, there seems to exist a common idea present in the different interpretations of QM, a common unspoken agreement between physicists and philosophers of physics regarding a basic metaphysical presupposition, previous to all the different ``ways of talking'' presented by each narrative. There is something that appears as indisputable, irrespectively of the choice of any interpretation, namely, that QM talks about a microscopic realm in which there are ``small'' quantum particles. 

As we have discussed elsewhere \cite{deRonde16b}, the present situation can only be understood when considering the deep influence of the empirical-positivist trend of thought which during the 20th Century, produced a major re-foundation of the meaning and understanding of physics. Replacing the account of physics which had ruled ---since the Greeks--- for more than two millennia, this new foundation grounded the reference of science in the ``common sense'' observations made by agents and their intersubjective agreement. This extreme form of relativism went explicitly against the original Greek foundation of {\it theories} as a difficult yet possible kind of discourse that could in fact enable an understanding of \emph{physis}. Rudolph Carnap, Otto Neurath and Hans Hahn argued explicitly in their famous Vienna manifesto \cite{VC} that: ``In science there are no `depths'; there is surface everywhere: all experience forms a complex network, which cannot always be surveyed and, can often be grasped only in parts. Everything is accessible to man; and man is the measure of all things. Here is an affinity with the Sophists, not with the Platonists; with the Epicureans, not with the Pythagoreans; with all those who stand for earthly being and the here and now.''  In this context, metaphysics was understood as a dogmatic discourse ---detached from experience--- about unobservable entities. Science should focus in ``statements as they are made by empirical science; their meaning can be determined by logical analysis or, more precisely, through reduction to the simplest statements about the empirically given.'' Still, and regardless of the influence of positivism within the development of QM, the understanding of physics as a discipline making reference to {\it physis} (or reality) was still shared by several of the founding fathers of QM who repeatedly stressed the direct link of their own ideas to those of the ancient Greeks. As remarked by Heisenberg: 
\begin{quotation}
\noindent {\small  ```Understanding' probably means nothing more than having whatever ideas and concepts are needed to recognize that a great many different phenomena are part of coherent whole. Our mind becomes less puzzled once we have recognized that a special, apparently confused situation is merely a special case of something wider, that as a result it can be formulated much more simply. The reduction of a colorful variety of phenomena to a general and simple principle, or, as the Greeks would have put it, the reduction of the many to the one, is precisely what we mean by `understanding'. The ability to predict is often the consequence of understanding, of having the right concepts, but is not identical with `understanding'.'' \cite[p. 63]{Heis71}}
\end{quotation}
From this viewpoint, theory comes first, and observation is secondary. As Einstein told Heisenberg, it is only the theory which tells you what can be observed. And in this particular respect, physical concepts play an essential role since they provide the possibility of understanding, of thinking in representational terms, even of making observations meaningful (this is, of putting them in relation to \emph{physis} thanks to the theory). 
\begin{quotation}
\noindent {\small``From Hume Kant had learned that there are concepts (as, for example, that of causal connection), which play a dominating role in our thinking, and which, nevertheless, can not be deduced by means of a logical process from the empirically given (a fact which several empiricists recognize, it is true, but seem always again to forget). What justifies the use of such concepts? Suppose he had replied in this sense: Thinking is necessary in order to understand the empirically given, {\it and concepts and `categories' are necessary as indispensable elements of thinking.}'' \cite[p. 678]{Einstein65} (emphasis in the original)}
\end{quotation} 
From a representational realist standpoint, there is no reference in physical theories to empirical subjects (or agents), to measurements and observations. Measurements play no role within the representation provided in formal-conceptual terms by a physical theory. In fact, theory construction is grounded on the mathematical invariance of the formalism which, in turn, provides the objectivity conditions required in order to detach itself from any particular reference frame or even observation made by an empirical subject \cite{deRondeMassri17}. As Einstein made the point, the moon must have a position whether or not we choose to observe it. And this fact is reflected theoretically, not only in the mathematical structure of the theory but also in the conceptual framework. Physical concepts are not isolated independent elements making reference to `things', they are part of an interrelated wholeness which allows the possibility of knowledge, beyond our individual perspectives, about the physical world. This does not mean a direct one-to-one correspondence reference between each concept and reality itself, but the constitution of a level in which observation and measurement are not the fundamental conditions to describe a state of affairs (they are, quite differently, a methodological tool, a constructed way of testing the adequacy of a physical theory).

Scientists do not make promises about the future possibilities of a theory, they do not talk about irrepresentable situations, they do not care about marketing, publishing or prizes, scientists produce a discourse which provides understanding about phenomena in objective (subject-independent) terms through the constitution of theoretical formal-conceptual {\it moments of unity}. Unscientific ways of talking, regardless of the fact they might be performed by communities of well established scientists, are not part of scientific discourse. A scientific discourse is not something relative to a community of people, it is a particular kind of discourse which has been prodigiously developed since the Greeks and for more than two millennia, with specific characteristics, attentions and difficulties that allows us to go beyond our mere subjective perspectives, individual opinions and provide an understanding of {\it physis}. Something, let us remark again, which does not mean to have a one-to-one correspondence relation between each concept as describing a \textit{thing} as it is \textit{in itself }---an idea used by postmodern trends of thought to caricaturize realism. 

From a representational realist standpoint, QM should be regarded as a proto-theory which, mainly due to the lack of a conceptual framework which is adequate to explain the mathematical formalism and the experience operationally predicted in the lab, is still incapable of expressing consistently what it talks about. As Heisenberg made the point \cite[p. 264]{Heis73}: ``The history of physics is not only a sequence of experimental discoveries and observations, followed by their mathematical description; it is also a history of concepts. For an understanding of the phenomena the first condition is the introduction of adequate concepts. Only with the help of correct concepts can we really know what has been observed.'' In this respect, the need to provide an {\it anschaulicht} (intuitive) access of the phenomena involved in QM is essential for an understanding of the theory. This is a necessary requirement for any meaningful scientific discourse about the theory of quanta which attempts to provide intelligibility. Something which ---at least, at the conceptual level--- non-individuals do not seem able to provide. In the following section, we attempt to show that there is a path grounded on the mathematical formalism of the theory itself which, going from the analysis of mathematical invariance into that of conceptual objectivity, might allows us to provide an {\it anschaulicht} representation of the theory of quanta, and in turn, an account of quantum individuality. One that goes beyond objects and particle metaphysics.

\section{Beyond Particle Metaphysics: Powers with Definite Potentia}

In \cite{Arenhart19}, Arenhart argued that: ``It seems that no metaphysician, in any tradition, would deny that our current scientific theories provide evidence to the claim that there are electrons, protons, spacetime, and so on.'' We disagree on this point. As analyzed from the perspective of the tradition we have been outlining ---a tradition which is certainly not marginal in the history of Western thought---, QM is not forced to make reference to electrons and protons conceived as elementary particles. The present discourse applied by quantum physicists and philosophers of physics which makes reference to quantum particles, quantum jumps and the like, is essentially meaningless. It provides no conceptual intelligibility whatsoever. To talk about electrons or protons without an understanding of what they are is just applying dogmatically degraded forms of atomist intuitions to a theory which simply does not seem to be talking about ``small objects'' living in space-time. As Heisenberg \cite{Boku04} remarked: ``the inappropriate transfer of classical concepts and ideas to the problems of atomic structure. . . . The program of quantum mechanics must therefore be first of all to free itself from these intuitive [anschaulichen] pictures, and, in place of the laws of classical kinematics and mechanics used up until now, to establish simple relations between experimentally given quantities.'' Together with Heisenberg \cite{Boku04}, we agree that ``the transition in science from previously investigated fields of experience to new ones will never consist simply of the application of already known laws to these new fields. On the contrary, a really new field of experience will always lead to the crystallization of a new system of scientific concepts and laws''. It is only once we recognize the depth of this representational and linguistic problem, that we will be actually able to advance through the secure path of scientific conceptual (metaphysical) development. We might remark in this respect the serious warning by Einstein: 
\begin{quotation}
\noindent {\small ``Concepts that have proven useful in ordering things easily achieve such an authority over us that we forget their earthly origins and accept them as unalterable givens. Thus they come to be stamped as `necessities of thought,' `a priori givens,' etc. The path of scientific advance is often made impossible for a long time  through such errors.'' \cite{Howard10}}
\end{quotation}
A first step to advance in the development of new conceptual forms is to abandon completely the reference to undefined ``elementary particles'' or ``quantum objects''. Taking as a standpoint the representational realist account of theories, the logos categorical approach to QM has attempted to develop a new non-classical conceptual framework which allows to restore an objective representation of the mathematical formalism. Our guide has been the orthodox mathematical formalism and, in particular, its invariant character. From this standpoint of analysis, we acknowledge the Born rule as the main operational-invariant expression of the theory, which gives us the quantities that remain the same, regardless of the perspective, arrangement or context. Taking this into consideration (and the fact that these invariant quantities are not binary) we noticed it was necessary to reach a new generalized definition of what could be considered to be an element of physical reality in QM (see \cite{deRonde16a}):

\smallskip
\smallskip

\noindent {\it {\bf Generalized Element of Physical Reality:} If we can predict in any way (i.e., both probabilistically or with certainty) the value of a physical quantity, then there exists an element of reality corresponding to that quantity.}

\smallskip
\smallskip
 
\noindent This definition, which goes beyond the {\it binary} reference to properties constituting systems, implies an intensive understanding of physical reality. What is intensive (and not only what is binary) can be considered as physically real. Taking as a standpoint Born's rule as an intensive quantification of projection operators we have advanced into the consideration of the notion of {\it power} with definite {\it potentia} (see for a detailed analysis \cite{deRonde19}). Those real intensive quantities determine the value (potentia) of powers, as physical elements that ---contrary to particles--- need not be redirected to binary values in order to be considered meaningful. In turn, our logos approach has allowed us to bypass the Kochen-Specker theorem and provide a {\it Global Intensive Valuation} to all projection operators independently of the context of inquiry \cite{deRondeMassri18a}. In this way, our approach provides an objective representation in which a lab is understood, not as an {\it Actual State of Affairs} described by a set of particles with well defined properties constituting objects in space-time nor as electromagnetic interacting fields, but instead, as a multiplicity of powers with definite potentia, all of them interacting and conforming a {\it Potential State of Affairs}. This might be the most important distinction between classical physics and QM, namely, that while the first represents reality in terms of a {\it binary} conceptual and mathematical representation, the latter makes explicit use of an {\it intensive} representation in which our binary Boolean logic and our classical concepts simply break down. 

Intuitively, we can picture a Potential State of Affairs (PSA) as a table which goes from the powers (projection operators) to the invariant number which quantifies their potentia through the Born rule,
\[
\Psi:\mathcal{G}(\mathcal{H})\rightarrow[0,1],\quad
\Psi:
\left\{
\begin{array}{rcl}
P_1 &\rightarrow &p_1\\
P_2 &\rightarrow &p_2\\
P_3 &\rightarrow &p_3\\
  &\vdots&
\end{array}
\right.
\]

\noindent Notice that our mathematical representation is objective in the sense that it relates, in a coherent manner and without internal contradictions, the multiple contexts (or aggregates) and the whole PSA. Contrary to the contextual (relativist) Bohrian ``complementarity solution'', there is in this case no need of a (subjective) choice of a particular context in order to define the ``physically real'' state of affairs (i.e., the definite binary values of properties). The state of affairs is described completely by the whole graph (or $\Psi$), and the contexts bear an invariant (objective) existence independently of any (subjective) choice. Let us remark that `objective' is understood as providing the conditions of a theoretical representation in which all subjects, observations and measurements are {\it detached} from the course of events. 

The notion of {\it power} presented in the logos approach, contrary to systems constituted by binary properties with an actual existence, has to be defined in intensive terms. A power is always quantified in terms of a {\it potentia} which measures its strength. The way to compute the potentia of each power is via the Born rule. Unlike classical systems, powers have a potential existence which must be defined in different categorical terms. Due to their invariant character, both powers and potentia can be regarded as possessing an objective existence, independent of contexts or measurements. In this way, these notions allow us to escape the widespread Bohrian dogma according to which we must simply accept that measurements in QM have a special status.\footnote{It is commonly argued that when we measure in QM, we always influence the quantum object under study ---which is just another way of making reference to the famous ``collapse'' of the quantum wave function. This idea, mainly due to Bohr's account of QM, implies that subjects define reality in an explicit manner; or as Bohr himself used to say: ``that in the great drama of [quantum] existence we are not only spectators but also actors.''} Kochen-Specker type contextuality becomes in our scheme a purely epistemic feature of the theory, one that deals with the {\it epistemic incompatibility} of quantum experiments, and not with the {\it ontic incompatibility} of quantum existents (see \cite{deRonde20, deRondeMassri18a}). The Born rule is not anymore understood in epistemic terms, as making reference to the probability of obtaining measurement outcomes. Instead, Born's rule is now conceived as a way of computing an objective feature of the (potential) state of affairs represented by QM (see for a detailed discussion \cite{deRonde16a, deRondeFreytesSergioli19}). The rule provides a definite value of the potentia of each power. Measurements become in this case only part of an epistemic procedure which collects statistical information about a (potential) state of affairs described in intensive terms. Consequently, there is no collapse, no real physical process taking place when an agent observes the result of a measurement process. The finding of an outcome relates then to a phenomena contained within the theoretical representation of the theory itself \cite{deRondeMassri19b}. All this work points in the direction of finding a consistent representation in order to address and understand the experience discussed by the theory of quanta.

\section{Final Remarks Regarding Quantum Individuality}

Our approach requires a consistent definition of quantum individual in both mathematical and conceptual terms. We agree with Ot\'avio Bueno\footnote{Discussins during the {\it V International Workshop on Foundations of Quantum Mechanics and Quantum Information} held in Buenos Aires, 13-14 March 2019. URL: https://quantuminternationalnet.com/Workshops.} that in order to develop a consistent notion which does not ground itself in atomism, it might be necessary to escape classical set theory right from the start. In this respect, category theory and, in particular, the notion of {\it aggregate} might provide us with a line of research that allows us to configure a completely new (non-classical) notion of quantum individuality. And we also agree with Arenhart and Kruse that quasi set theory and non-reflexive logics might also provide an interesting standpoint of analysis. However, as we have argued, the intelligibility regarding quantum individuality must be provided also in conceptual metaphysical terms. In turn, such intelligibility must be able to produce a consistent intuitive understanding of the state of affairs represented by the theory of quanta. We are aware that, besides from entering this interesting discussion Arenhart, Krause and Bueno have been developing, we should be able to give a coherent and extensive account of individuality in QM from the standpoint we have outlined. This is the object of ongoing work, and will be surely the subject of coming publications.


\end{document}